\documentclass[sn-nature]{sn-jnl}

\usepackage{graphicx}%
\usepackage{multirow}%
\usepackage{amsmath,amssymb,amsfonts}%
\usepackage{amsthm}%
\usepackage{mathrsfs}%
\usepackage[title]{appendix}%
\usepackage{xcolor}%
\usepackage{textcomp}%
\usepackage{manyfoot}%
\usepackage{booktabs}%
\usepackage{algorithm}%
\usepackage{algorithmicx}%
\usepackage{algpseudocode}%
\usepackage{listings}%

\usepackage{bm}
\usepackage{lmodern}

\begin{document}

\title[Article Title]{Full Polarization Control of Photons with Evanescent Wave Coupling in the Ultra Subwavelength Gap of Photonic Molecules}


\author[1,2]{\fnm{Rui} \sur{Zhu}}
\equalcont{These authors contributed equally to this work.}
\author[1,2]{\fnm{Chenjiang} \sur{Qian}}
\equalcont{These authors contributed equally to this work.}
\author[1,2]{\fnm{Shan} \sur{Xiao}}
\equalcont{These authors contributed equally to this work.}
\author[3]{\fnm{Jingnan} \sur{Yang}}
\author[1,2]{\fnm{Sai} \sur{Yan}}
\author[4]{\fnm{Hanqing} \sur{Liu}}
\author[4]{\fnm{Deyan} \sur{Dai}}
\author[3]{\fnm{Hancong} \sur{Li}}
\author[3]{\fnm{Longlong} \sur{Yang}}
\author[3]{\fnm{Xiqing} \sur{Chen}}
\author[1,2]{\fnm{Yu} \sur{Yuan}}
\author[1,2]{\fnm{Danjie} \sur{Dai}}
\author*[1,2]{\fnm{Zhanchun} \sur{Zuo}}\email{zczuo@iphy.ac.cn}
\author[4]{\fnm{Haiqiao} \sur{Ni}}
\author[4]{\fnm{Zhichuan} \sur{Niu}}
\author*[1,2]{\fnm{Can} \sur{Wang}}\email{canwang@iphy.ac.cn}
\author[1,2]{\fnm{Kuijuan} \sur{Jin}}
\author[3]{\fnm{Qihuang} \sur{Gong}}
\author*[3,5,6]{\fnm{Xiulai} \sur{Xu}}\email{xlxu@pku.edu.cn}

\affil[1]{Beijing National Laboratory for Condensed Matter Physics, Institute of Physics, Chinese Academy of Sciences, Beijing 100190, China}
\affil[2]{School of Physical Sciences, University of Chinese Academy of Sciences, Beijing 100049, China}
\affil[3]{State Key Laboratory for Mesoscopic Physics and Frontiers Science Center for Nano-optoelectronics, School of Physics, Peking University, 100871 Beijing, China}
\affil[4]{State Key Laboratory of Superlattices and Microstructures, Institute of Semiconductors Chinese Academy of Sciences, Beijing 100083, China}
\affil[5]{Peking University Yangtze Delta Institute of Optoelectronics, Nantong, Jiangsu 226010, China}
\affil[6]{Collaborative Innovation Center of Extreme Optics, Shanxi University, Taiyuan, Shanxi 030006, China}

\abstract{
Polarization of photons plays a key role in quantum optics and light-matter interactions, however, it is difficult to control in nanosystems since the eigenstate of a nanophotonic cavity is usually fixed and linearly polarized.
Here we reveal polarization control of photons using photonic molecules (PMs) that host supermodes of two coupled nanobeam cavities.
In contrast to conventional PMs in a 2D photonic crystal slab, for the two 1D photonic crystal nanobeam cavities the shift and gap between them can be tuned continuously.
With an ultra subwavelength gap, the coupling between the two cavities is dominated by the evanescent wave coupling in the surrounding environment, rather not the emission wave coupling for conventional PMs.
As such, non-Hermiticity of the system becomes pronounced, and the supermodes consist of a non-trivial phase difference between bare eigenstates that supports elliptical polarization.
We observe that both the polarization degree and polarization angle of the antisymmetric mode strongly depend on the shift and gap between the two cavities, exhibiting polarization states from linear to circular.
This full polarization control indicates great potential of PMs in quantum optical devices and spin-resolved cavity quantum electrodynamics.
}




\maketitle

\section{Introduction}\label{sec1}

Nanobeam cavity is one-dimensional (1D) dielectric beam with photonic crystal nanostructures that confine photons at resonant energy.
The cavity mode has advantages including a subwavelength mode volume,  high quality (Q) factor, and small device footprint \cite{10.1103/PhysRevLett.118.223605,10.1038/ncomms6580,10.1063/1.3568897}.
The design flexibility of the nanobeam allows for \textit{in situ} integration and coupling to active emitters such as quantum dots \cite{10.1103/PhysRevLett.114.143603,10.1103/PhysRevB.86.075314,10.1002/lpor.202100009} and 2D materials \cite{10.1103/PhysRevLett.128.237403,10.1021/acsphotonics.8b00036}, resulting in exceptional performances and control over the light-matter interactions.
As such, the nanobeam cavity has been widely applied to manipulate the local optical field for nanophotonic devices such as low-threshold lasing \cite{10.1038/ncomms3822,10.1021/nl504432d,10.1038/nnano.2017.128,10.1126/sciadv.adk6359}, ultra-precise sensing \cite{10.1021/nn4050547,10.1021/acsphotonics.5b00602,10.1109/JPHOT.2016.2536942}, optomechanical coupling \cite{10.1038/nature08524,10.1364/OPTICA.6.000213,10.1038/s41567-019-0673-7}, and quantum interfaces between different degrees of freedom \cite{10.1038/s41586-020-3038-6,10.1038/s41566-020-0609-x,PhysRevLett.130.126901}.

Polarization of photons is an inherent property of the cavity mode and plays key roles in photon generation, photon detection, and light-matter interactions \cite{10.1038/s42254-021-00398-z,10.1103/PhysRevLett.98.063601}.
The manipulation of resonant energy and mode profile of a single nanocavity has been widely studied \cite{10.1007/s11433-015-5724-1,10.1364/JOSAB.398574,10.1364/OME.415128}, but however, the polarization of photons emitted from the cavity mode is usually linear due to the break of rotational symmetry \cite{10.1063/1.2748310}.
Full polarization control in nanophotonic devices has been attempted by using passive polarization filters \cite{10.1021/acsphotonics.8b00522,10.1038/s41377-019-0184-4,10.1002/adom.201900129,10.1038/s41467-022-31726-1,10.1002/lpor.202300501} or by coupling them to nanostructures \cite{10.1038/ncomms7695,PhysRevB.101.245308}.
In contrast, the fabrication of a cavity with fully polarized eigenstates would be a direct and efficient way to control the interaction with integrated active emitters.
The key to polarization control is non-Hermiticity, which is responsible for complex supermodes, because elliptical polarization is the supermode of linear polarizations with an imaginary phase shift.
For example, circular polarized mode has been reported for a single cavity with non-Hermitian coupling between two degenerate modes \cite{PhysRevResearch.3.043096}.

Compared to a single cavity, photonic molecules (PMs) host not only supermodes of coupled cavities but also the high controllability arising from the freedom of separation between the two cavities \cite{PhysRevLett.81.2582,10.1002/lpor.200910001,10.1364/OE.21.016934,10.1364/OE.23.009211,10.1038/s41566-018-0317-y,10.1364/AOP.376739}.
Chalcraft et al. have investigated the PM in a 2D photonic crystal slab and found that the polarization angle of the antisymmetric (AS) mode strongly depends on the separation between cavities \cite{10.1364/OE.19.005670}.
However in the 2D slab, cavities are located at the lattice points of the photonic crystal, and thereby, the separation cannot be continuously tuned towards the ultra subwavelength regime.
The large separation results in symmetric and Hermitian coupling, and the polarization of supermodes retains linear.

In this work, we reveal the full polarization control of photons using PM consisting of two identical nanobeam cavities.
The photonic molecule is controlled by the gap between the two nanobeams and the shift of cavity centers along the nanobeam.
The gap and shift are tuned continuously for high controllability and unique features arising from the large coupling strength \cite{10.1103/PhysRevApplied.13.044041,10.1038/s41467-023-41127-7}.
As the gap decreases from the wavelength scale to the ultra subwavelength scale, the PM turns from far-field wave coupling to evanescent wave coupling, which provides the basis for non-Hermiticity and exceptional effects \cite{10.1364/AO.50.006272,10.1126/science.aar7709,10.1038/s42005-023-01508-2}.
This non-Hermiticity is demonstrated by the observation of supermodes beyond the prediction of coupled mode theory in the Hermitian regime, such as the non-trivial phase shift between the symmetric (S) and AS modes.
Furthermore, we observe that for the AS mode, both the angle and degree of polarization strongly depend on the gap and shift.
The full polarization control from linear to circular is consistent with the evanescent wave coupling, and is further strengthened by the control case with far-field wave coupling in which no polarization control is observed.

\section{Results}\label{sec2}

\begin{figure}
    \centering
	\includegraphics[width=1\linewidth]{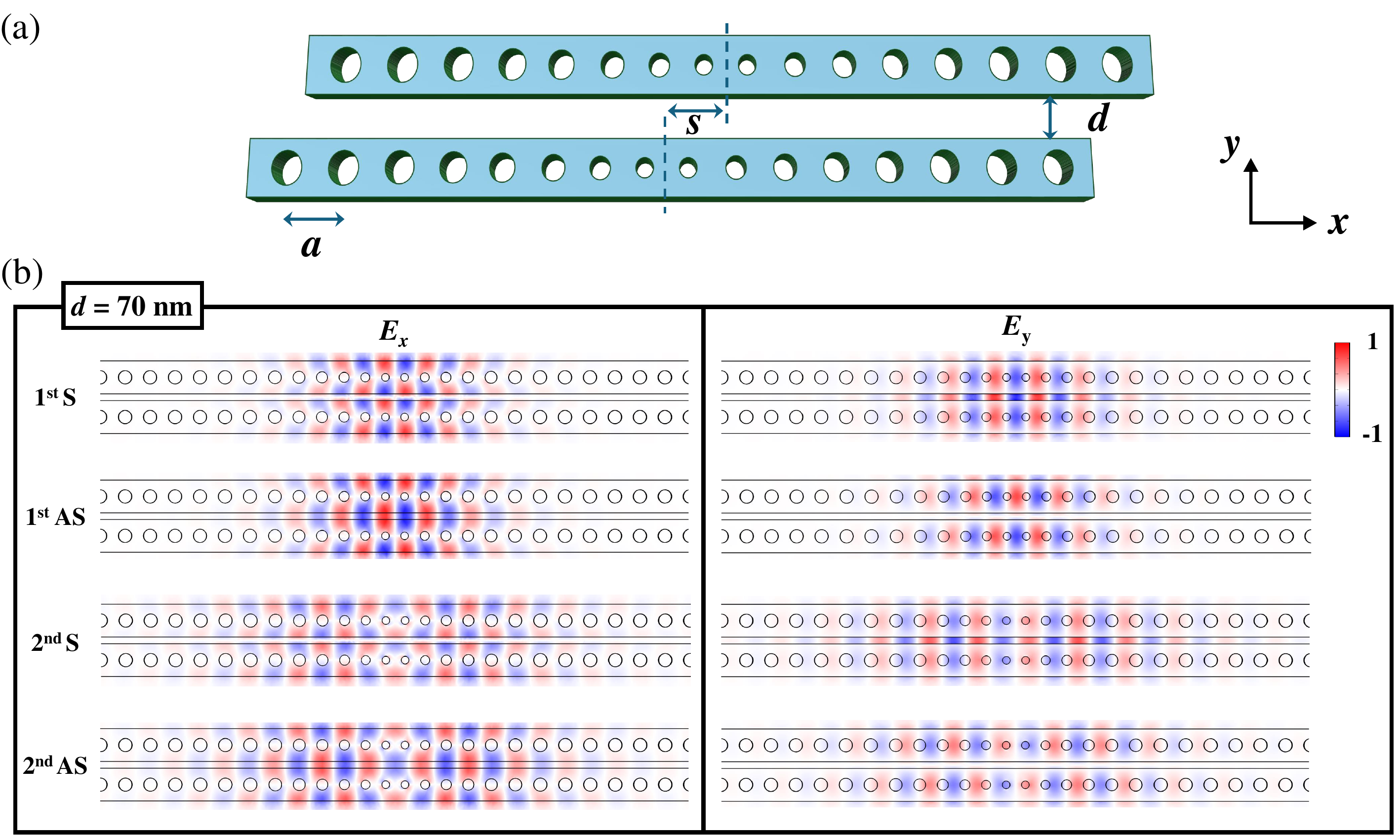}
	\caption{
        (a) Schematic of a PM that consists of two nanobeam cavities.
        The coupling between two cavities is controlled by the gap $d$ and shift $s$.
        (b) Normalized profile of the electric field ($E_{x}$ at left and $E_{y}$ at right) for the S and AS mode with $d=70$ nm, identified by the phase difference 0 and $\pi$ in the two cavities.
        }
	\label{p1}
\end{figure}

We study the PM consisting of two GaAs nanobeam cavities as schematically depicted in Fig. \ref{p1} (a).
The nanobeam has a thickness of 150 nm and a width of 340 nm.
The 1D photonic crystal nanoholes have the lattice constant $\textit{a}$ and radius $\textit{r}$, slowly chirped from the center to the two ends ($\textit{a}$: 260 to 200 nm, $\textit{r}$: 68.8 to 40.3 nm) for a smooth confinement of photons \cite{10.1063/1.3107263}.
Such cavities typically exhibit a high Q factor $> 10000$ and linear $y$ polarization for the first fundamental cavity mode.
In experiment, we fabricate the PM in a GaAs membrane embedded with InGaAs quantum dots (QDs) grown by molecular beam epitaxy.
The ensemble QDs have a high density, exhibiting broad emission with a wavelength range from 950 to 1100 nm, as light sources to excite the supermodes.
Details of the design and fabrication method can be found in the Supplementary materials.

The coupling between two nanobeam cavities is controlled by the gap $d$ and shift $s$, as depicted in Fig. \ref{p1} (a).
We resolve the eigenstates of PM using 3D finite element method.
Typical results with $d=70$ nm and $s=0$ nm are presented in Fig. \ref{p1} (b).
As shown, the supermodes include the S and AS modes, identified by whether the electric field profiles in the two cavities have the same phase or a $\pi$-phase difference.
Conventionally, the coupling in PM is described by the coupled mode theory in the Hermitian regime as $\dot{a}_{1,2}=-i\omega_0 a_{1,2}-\left(\gamma_0/2\right) a_{1,2}-iga_{2,1}$, where $a_{1,2}$ is the field amplitude of the first and second cavity, $\omega_0$ ($\gamma_0$) is the bare energy (linewidth) of a single cavity, and $g$ is the coupling strength between two cavities \cite{10.1364/AOP.376739,10.1515/nanoph-2023-0347}.
Hence, the S and AS mode are linear superpositions as $\vert 1 \rangle \pm \vert 2 \rangle$ with the trivial 0 and $\pi$ phase shift, in which $\vert 1 \rangle$ and $\vert 2 \rangle$ are the bare eigenstates of each single cavity, and their eigenenergies are $\omega_0\pm g$ with a symmetric splitting.

\begin{figure}
    \centering
    \includegraphics[width=0.66\linewidth]{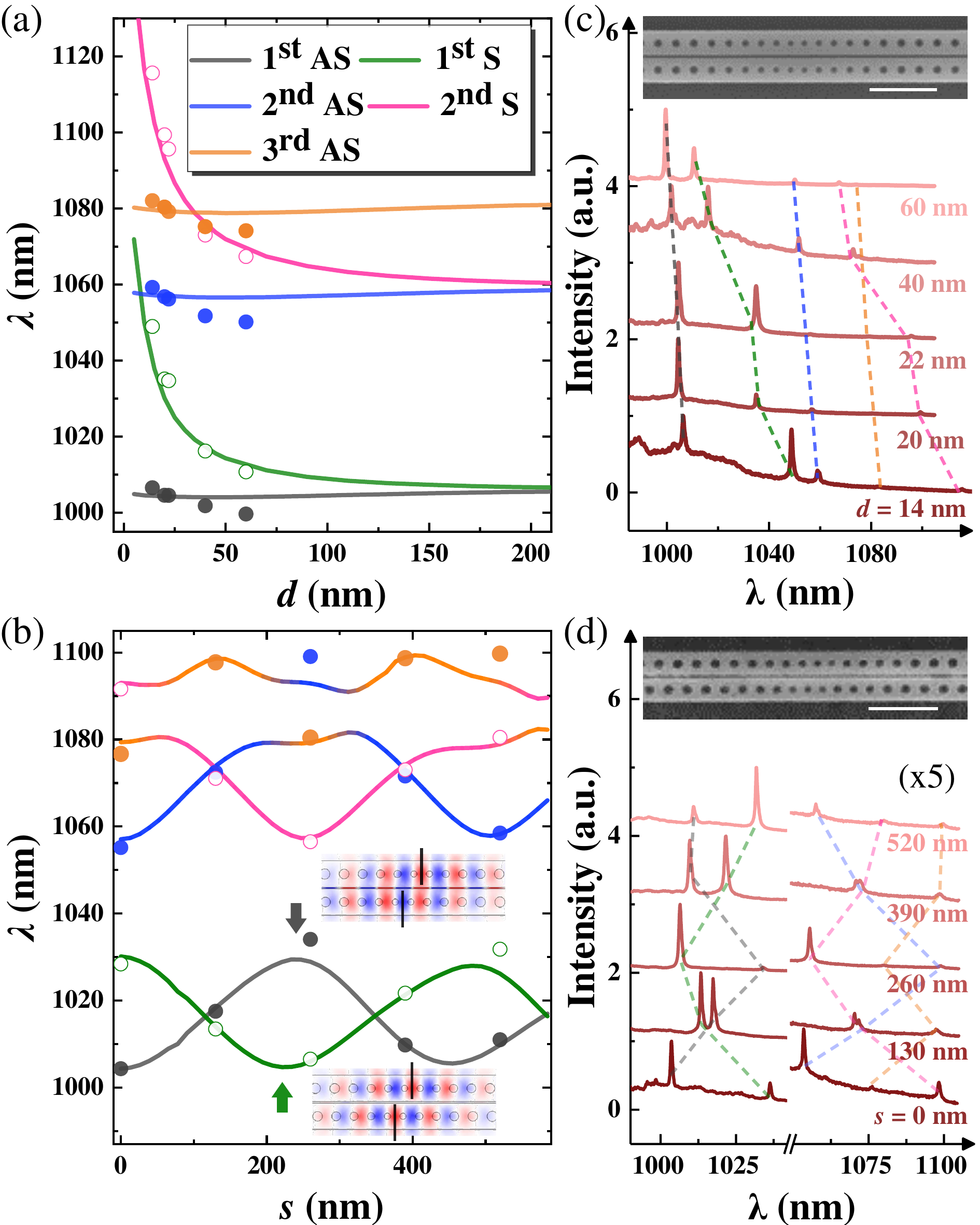}
	\caption{
        Supermodes as a function of (a) gap $d$ and (b) shift $s$.
        In (b) the arrows and field profiles of $E_{y}$ depict the maximum wavelength of the S mode at $s=240$ nm and the minimum of the AS mode at $s=225$ nm.
        Black lines in the field profiles denote the cavity center.
        (c)(d) Experimental PL spectra corresponding to data points in (a)(b).
        Sharp peaks represent the supermodes.
        Scanning electron microscope images show two fabricated PMs, one without and one with a shift of 130 nm, maintaining a gap of 20 nm.
        The scale bar is 1 $\mu$m.
        }
	\label{p2}
\end{figure}

However, the Hermitian coupling is only valid for a small coupling strength $g$ with a large gap, and as the gap decreases non-trivial effects arise.
In Fig. \ref{p2} (a) we present results with varying gap $d$ from 0 to 200 nm and no shift $s=0$ nm.
As shown, when $d>120$ nm, a symmetric splitting is observed for both the first and second mode.
The second mode exhibits a larger splitting because the field distribution extends more into the surrounding environment, which means a larger spatial overlap and a larger coupling strength.
In contrast, when $d<120$ nm, the S modes red shift rapidly, whilst the AS modes rarely shift.
Such asymmetric behavior can be explained by the second order nonlinearity or self coupling  \cite{10.1063/1.3176442,10.1002/lpor.201100017,10.1364/OE.14.001208}, which demonstrates that the Hermitian coupling is not enough to describe the PM with the small gap $d$.

To further explore the asymmetry in PM, we investigate the supermodes by controlling the shift $s$, as presented in Fig. \ref{p2} (b). 
The first S and AS mode shift quasi periodically along with a decay envelope as $s$ increases.
The periodicity is because the field profile of a single cavity includes positive and negative regions with quasi periodicity; thus, the overlap and anti-overlap switch as $s$ increases \cite{10.1038/s41598-020-79915-6}.
The decay envelope corresponds to the envelope of the field profile in a single cavity that is confined at the cavity center.
Typically the second supermodes should follow the quasi periodicity similar to the first order.
But here the second S and AS modes have such a large splitting that their energies cross the third AS mode, resulting in additional coupling, as shown by the blue, pink, and orange lines in Fig. \ref{p2} (b), which are also observed in experiment.
We excite the QDs in PMs with a 532 nm cw laser and collect the photoluminescence (PL) spectra as presented in Fig. \ref{p2} (c)(d), corresponding to the cases of varying $d$ and $s$, respectively.
The broad baseline emission is from the ensemble of QDs, and the sharp peaks arise from the supermodes of PM.
In Fig. \ref{p2} (d) the intensity of high order modes from 1050 to 1100 nm is relatively small; thus, they are multiplied by five times for clarity.
The wavelength (energy) of supermodes extracted from the experiment is presented by the data points in Fig. \ref{p2} (a)(b), exhibiting a good agreement with the theoretical calculations.

\begin{figure}
    \centering
    \includegraphics[width=0.66\linewidth]{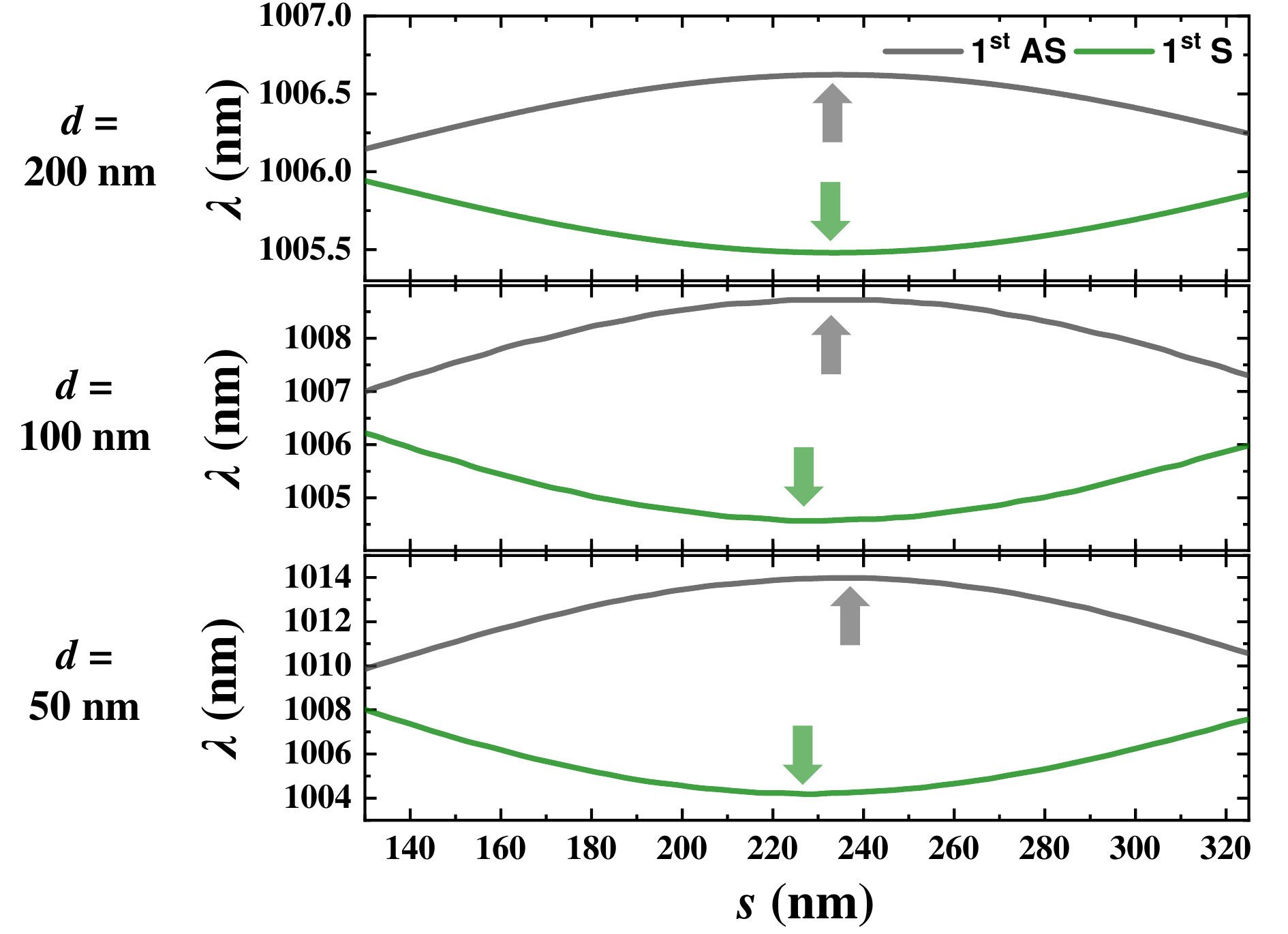}
    \caption{
        Supermodes as a function of shift $s$ when $d$ = 50, 100, and 200 nm.
        The additional phase differences of 8, 4, and 1 nm between the S and AS modes are observed for the three cases, respectively, as denoted by the arrows.
    }
    \label{p3}
\end{figure}

Moreover, we observe that in Fig. \ref{p2} (b), the antinodes of the S and AS modes exhibit a phase shifts as denoted by the gray and green arrows.
The antinodes of the first AS mode (gray) are at $s=$ 0, 240, and 452 nm, while the antinodes of the first S mode (green) are at 0, 225, and 487 nm.
The second and third antinodes have phase shift of 15 and 35 nm, respectively.
For the Hermitian coupling, the S and AS modes are linear superposition of the two bare single cavities with the phase of 0 and $\pi$; thus, the minimum wavelength position (anti-overlap of the field) of one mode should always correspond to the maximum wavelength position (overlap of the field) of the other.
In contrast, the additional phase shift in Fig. \ref{p2} (b) reveals the non-Hermitian features, i.e., the eigenstates are complex supermodes with imaginary components arsing from the evanescent wave coupling, rather not the trivial $\vert 1 \rangle \pm \vert 2 \rangle$ in the Hermitian regime \cite{10.1364/OE.22.012359}.
We additionally present this mismatch with the gap $d$ = 50, 100, and 200 nm in Fig. \ref{p3}.
The value of the phase shift for the antinodes at $\sim$ 230 nm is 8, 4, and 1 nm, respectively, clearly showing that the phase shift is suppressed as the gap $d$ increases.
Such non-trivial behavior agrees remarkably with the asymmetric splitting observed in Fig. \ref{p2} (a), supporting that with the ultra subwavelength $d$ the coupling between two cavities is dominated by the evanescent field.

\begin{figure}
    \centering
	\includegraphics[width=0.66\linewidth]{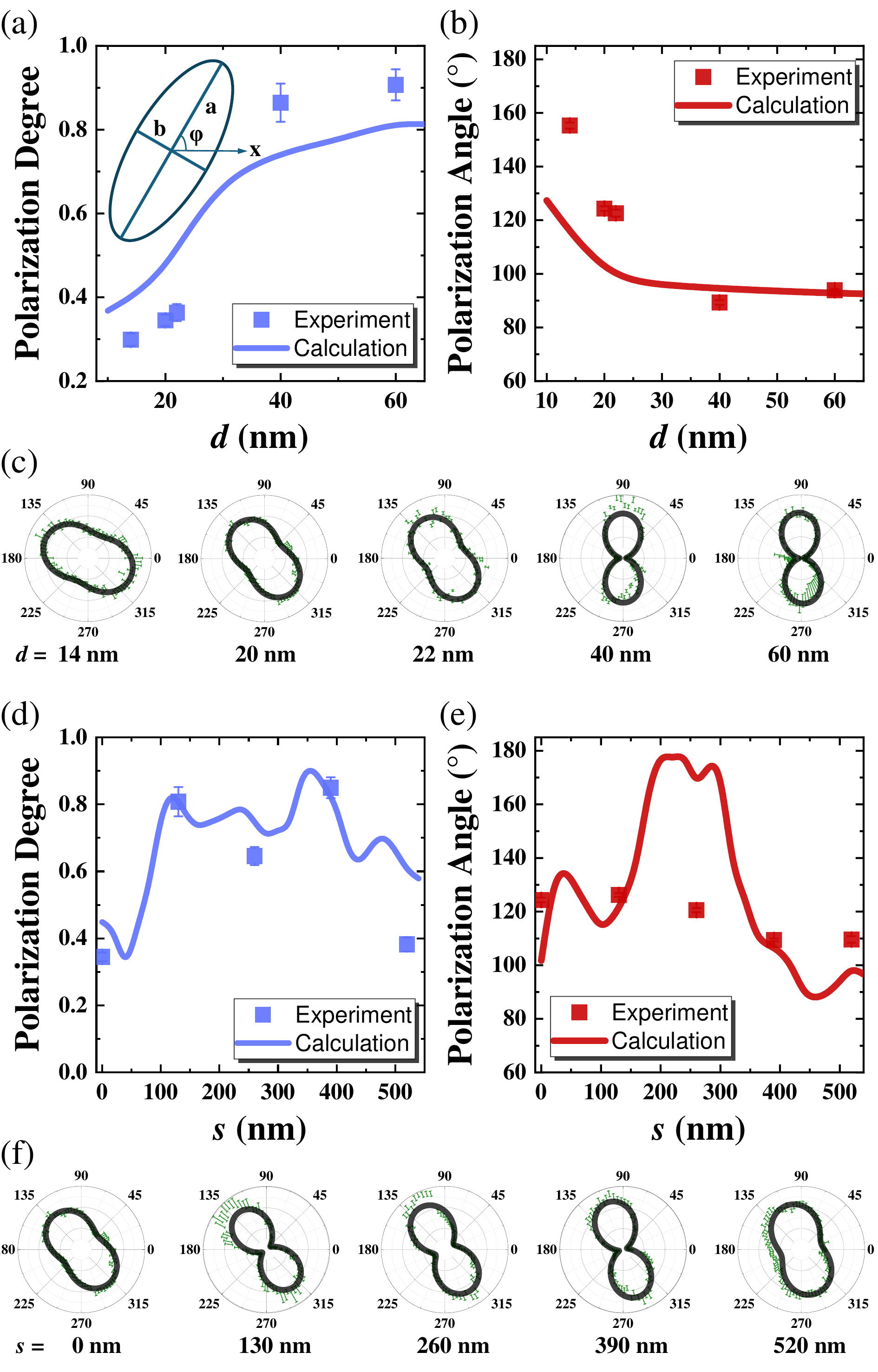}
	\caption{
        Experimentally measured polarization of photons emitted from the first AS mode.
        (a) Degree and (b) angle of polarization extracted from (c) polarization resolved peak intensities with the dependence on gap $d$.
        (d)-(f) Same measurement but with a different shift $s$.
        Solid lines are the calculation results.
    }
	\label{p4}
\end{figure}

\begin{figure}
    \centering
    \includegraphics[width=0.66\linewidth]{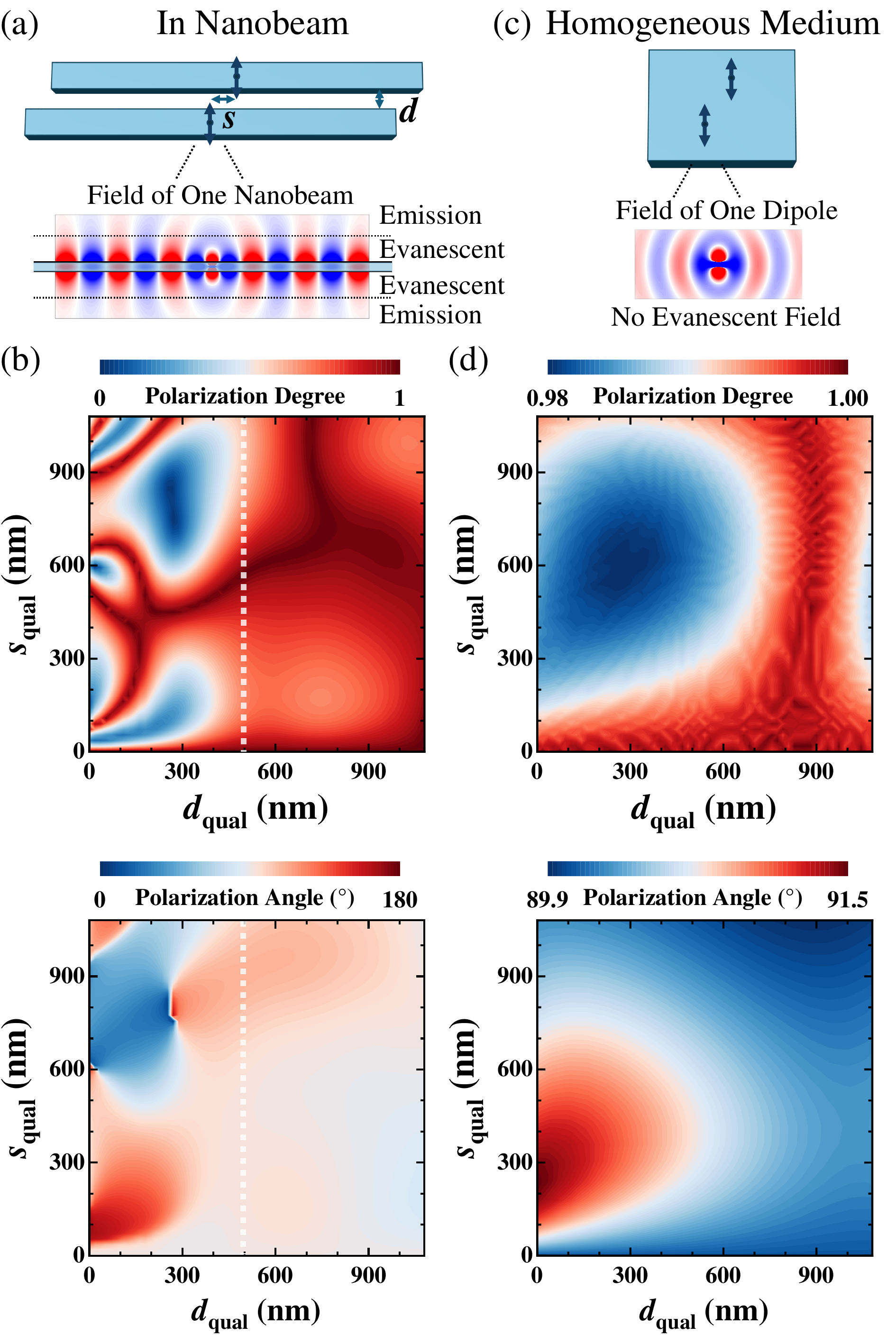}
    \caption{
        (a) Theoretical model with a $y$-polarized dipole in each nanobeam as an approximation of the $y$-polarized nanobeam cavity in Fig. \ref{p1}(a).
        Due to the total internal reflection of GaAs, evanescent field dominates near the nanobeam (within 500 nm), whilst the emission field dominates far away from the nanobeam.
        This essential feature is the same as that of the real nanobeam cavity in Fig. \ref{p1}(a).
        (b) Calculated polarization degree and angle as functions of gap $d_\mathrm{qual}$ and shift $s_\mathrm{qual}$.
        White dashed lines denote the threshold $d_\mathrm{qual}$ of 500 nm for the evanescent wave coupling.
        (c) Theoretical model with dipoles in a homogeneous medium.
        No evanescent field exists in this case.
        (d) Calculated polarization degree and angle for the case of homogeneous medium.
    }
    \label{p5}
\end{figure}

The evanescent wave enables the control of polarization by introducing a phase shift between the in-plane and out-of-plane electric field at the boundary of dielectrics \cite{10.1364/AO.50.006272}.
The polarization of cavity mode is related to the symmetry of the electric field \cite{10.1364/OE.15.017221,10.1063/1.2748310,10.1109/JQE.2002.1017597}.
In a single cavity, the $E_y$ of the first mode is symmetric in both the $x$ and $y$ axes, and the $E_x$ is anti-symmetric in both the $x$ and $y$ axes.
As such, the overall projection of the electric field only has the $y$-component while the $x$-component is zero, resulting in the polarization of the photon being linear in the $y$ direction.
The first S mode in Fig. \ref{p1} (b) inherits this symmetry, thus it is also $y$ polarized.
In contrast, in Fig. \ref{p1} (b), for the first AS mode $E_y$ is symmetric in the $x$ direction while anti-symmetric in the $y$ direction, and $E_x$ is anti-symmetric in the $x$ direction while symmetric in the $y$ direction.
In this case, the overall projection cannot be estimated from the features of symmetry but depends on the specific electric field distribution.
Therefore, the polarization of the AS mode is expected to be non-trivial and vary with the coupling.
Figs. \ref{p4} (a) and (b) show the degree and angle of polarization for the AS mode, respectively, in the case with $s=0$ nm and variable $d$.
The polarization ellipse is defined as shown in the inset of Fig. \ref{p4}(a) including the long axis $a$, the short axis $b$, and the angle of the long axis relative to the x direction $\phi$.
The polarization degree is defined by $\vert a-b \vert/\vert a+b \vert$, where 1 refers to linear polarization and 0 means circular polarization.
The polarization angle is defined by $\phi$ with values ranging from 0$^\circ$ to 180$^\circ$, in which 90$^\circ$ means $y$ polarized. 
Data points in Figs. \ref{p4} (a) and (b) are extracted from the polarization resolved PL intensities in Fig. \ref{p4} (c), and solid lines are the calculated results.
The experimental data generally follow the predictions of theoretical calculations and reveal a threshold value of $d=40$ nm in both the change of polarization degree and polarization angle.
For $d>40$ nm, the AS mode exhibits nearly linear polarization along the $y$ direction, e.g., at $d=$ 60 nm the polarization degree is 0.91, and the polarization angle is 93$^\circ$.
This linear polarization along the y direction is the same as the mode of a single cavity.
When $d<40$ nm, the polarization of the AS mode becomes elliptical, and the major axis turns towards the $x$ direction.
At $d=14$ nm, we achieve the smallest polarization degree of 0.29, which is close to circular polarization, and the largest polarization angle shifts to 155$^\circ$, which is nearly along the x direction.
In Figs. \ref{p4} (d) and (e) we present the degree and angle of polarization for the AS mode in the case with $d=20$ nm with variable $s$, extracted from the data in Fig. \ref{p4} (f).
The experimental $s$ dependence generally follows the prediction in theory.
The minor difference between the experiment and theoretical calculations mainly arises from the fluctuations in the gap $d$, i.e., here $d$ is designed to be 20 nm in calculation but varies by $\sim 5$ nm in experiment.
At such an ultra subwavelength gap, the fluctuation of $d$ results in the difference in experiment because the polarization is sensitive to $d$ as discussed next in Fig. \ref{p5}.

We next explore the polarization by varying both the gap and shift.
For brevity, we use a qualitative model to calculate the far-field polarization, as depicted in Fig. \ref{p5}(a).
A nanobeam with a $y$-polarized electric dipole is used to approximate the cavity because the single cavity mode is $y$-polarized.
The inset in Fig. \ref{p5}(a) shows the electric field $E_{y}$ of a single nanobeam embedded with a dipole.
Since GaAs has a high refractive index of 3.45, total internal reflection results in the domination of the evanescent field within the range of 500 nm away from the nanobeam.
This essential physics for the coupling in PM is the same as the real nanobeam shown in Fig. \ref{p1}(a), supporting the validity of our model.
Other details of the calculation are discussed in the supplementary materials.
In Fig. \ref{p5}(b), we present the degree and angle of polarization calculated with quasi-continuous gap $d_\mathrm{qual}$ and shift $s_\mathrm{qual}$.
A threshold of $d_\mathrm{qual}=500$ nm is observed.
The polarization is nearly $y$ polarized ($\phi \sim 90^\circ$) for $d_\mathrm{qual}>500$ nm, whilst it strongly depends on $d_\mathrm{qual}$ and $s_\mathrm{qual}$ for $d_\mathrm{qual}<500$ nm.
This threshold value agrees remarkably with the boundary of the evanescent field shown in Fig. \ref{p5}(a), and the threshold behavior is fully consistent with the real nanobeam cavity shown in Fig. \ref{p4}.
Moreover, for the evanescent wave coupling with $d_\mathrm{qual}<500$, the full polarization control from linear to circular (degree from 1 to 0) is achieved. 
The threshold $d_\mathrm{qual}$ of 500 nm, as shown in the theoretical calculation in Fig. \ref{p5}(b), is large compared to the threshold $d$ of 40 nm in experiment.
This is because the strength of the evanescent field is different in the two cases, as the model in Fig. \ref{p5}(a) is an approximation and not entirely the same as the nanobeam cavities shown in Fig. \ref{p1}(a). 

In Fig. \ref{p5}(c) we show the control case that the two dipoles are in a homogeneous medium.
The dipole emits in space without reflections, and thus no evanescent field exists.
In Fig. \ref{p5}(d), we present the degree and angle of polarization in this case.
As shown, the polarization degree changes from 1 to 0.98, and the polarization angle changes from 91.5$^\circ$ to 89.9$^\circ$.
Such a small  change in polarization is well explained by the lack of evanescent wave coupling.
The comparison between the nanobeam case and the homogeneous case in Fig. \ref{p5} further strengthens the polarization control in our PMs through the evanescent field in the ultra subwavelength gap.

The conventional passive filters for polarization control can only filter the photons that have already been generated by the emitter.
This means the generation and polarization control of photons are separated, and thereby, many photons are lost during the filtering.
Since the passive filter does not control the local optical field for the quantum emitter, it is independent of the light-matter interaction of the quantum emitter.
In contrast, in our sample the PM directly controls the polarization of the cavity eigenstate mode, and the emitters are embedded in the cavity.
This means we directly control the local optical field that couples to the emitter, indicating high efficiency in polarization control and the significant potential for applications in spin-resolved cavity quantum electrodynamics.
For example, the AS mode has stable resonant energy and controllable polarization with the varying gap.
The coupling of the AS mode to the embedded quantum emitter would pave the way for studying the polarization-dependent light-matter interactions, revealing the spin properties of the quantum emitter.

\section{Discussion}\label{sec3}

In summary, we demonstrate the full polarization control of photons using PMs in both theory and experiment.
The elliptical polarization arises from the non-Hermitian coupling through the evanescent field in the ultra subwavelength gap between two nanobeam cavities.
The threshold between evanescent wave coupling and far-field wave coupling is clearly observed when changing the gap.
These results indicate the great potential of PM in applications for spin-resolved cavity quantum electrodynamics, because the PM directly controls the local optical field that couples to the emitters embedded in the cavity.
The evanescent wave in our device also provides the basis for controlling the momentum and other degrees of freedom of photons \cite{10.1038/ncomms4300,10.1364/OPTICA.3.000118}.
In addition to the recently developed methods to control the separation in PMs using different degrees of freedom \cite{10.1103/PhysRevApplied.13.044041,10.1038/s41467-023-41127-7}, the non-trivial features indicate the great potential of the PMs in applications based on quantum optics and cavity quantum electrodynamics.

\backmatter

\bmhead{Materials and methods}
The sample with QDs in this work was grown by molecular-beam epitaxy method.
From top to bottom, the sample consists of three layers including a 150-nm-thickness GaAs layer embedded with InGaAs QDs, a 1-$\mu$m-thickness AlGaAs sacrifice layer for the underetching of PM, and a GaAs substrate.
Electron beam lithography was used to pattern the PM structures with a resist mask on the sample.
Then inductively coupled plasma was used to etch the unprotected GaAs regions below the patterned resist.
In the end, wet etching with HF solutions was used to remove the AlGaAs sacrificial layer below the PM structures. 
The PMs were mounted in a cryostat and cooled down to 4.2 K by liquid helium flow.
InGaAs QDs were excited by a cw laser with a wavelength of 532 nm, and the PL signals were collected by a spectrometer with InGaAs detectors.
The polarization dependence was resolved by filtering the PL signals through a rotating half-wave plate and a polarizer.
The theoretical results were calculated with 3D finite element method (FEM) and finite-difference time-domain (FDTD) methods.

\bmhead{Acknowledgements}
This work was supported by the National Key Research and Development Program of China (Grant No. 2021YFA1400700), the National Natural Science Foundation of China (Grants Nos. 62025507, 12494600, 12494601, 12494603, 11934019, 92250301, 22461142143, 62175254, 12174437 and 12204020).

\bmhead{Data Availability}
The datasets generated and analysed during the current study are available from the corresponding authors upon reasonable request.

\bmhead{Conflict of Interest}
Authors state no conflict of interest.

\bmhead{Contributions}
X. X., C. W., K. J., and Q. G. supervised the research project.
R. Z. and S. X. designed and fabricated the devices and conducted the optical measurements.
R. Z. and C. Q. performed the theoretical calculations.
J. Y., S. Y., H. L., L. Y., X. C., Y. Y., D. D., Z. Z., and X. X. helped the experiments and data analysis.
H. L., D. D., H. N., and Z. N. grown the substrate with QDs. 
All the authors have contributed to writing the manuscript.

\bmhead{Supplementary materials}
The supplementary material file includes the detailed methods for theory and experiments, as well as typical results of control cases.

\end{document}